\input harvmac
\newcount\figno
\figno=0
\def\fig#1#2#3{
\par\begingroup\parindent=0pt\leftskip=1cm\rightskip=1cm
\parindent=0pt
\baselineskip=11pt
\global\advance\figno by 1
\midinsert
\epsfxsize=#3
\centerline{\epsfbox{#2}}
\vskip 12pt
{\bf Fig. \the\figno:} #1\par
\endinsert\endgroup\par
}
\def\figlabel#1{\xdef#1{\the\figno}}
\def\encadremath#1{\vbox{\hrule\hbox{\vrule\kern8pt\vbox{\kern8pt
\hbox{$\displaystyle #1$}\kern8pt}
\kern8pt\vrule}\hrule}}

\overfullrule=0pt

\Title{\vbox{\baselineskip12pt
\hbox{hep-th/0008088}
\hbox{TIFR-TH/00-43}
\hbox{OHSTPY-HEP-T-00-012}
}}
{\vbox{\centerline{Giant Gravitons, BPS bounds and Noncommutativity}}}
\smallskip
\centerline{Sumit R. Das},
\smallskip
\centerline{{\it Tata Institute of Fundamental Research}}
\centerline{\it Homi Bhabha Road, Mumbai 400 005, INDIA}
\bigskip
\centerline{Antal Jevicki}
\smallskip
\centerline{{\it Department of Physics, Brown University}}
\centerline{\it Providence, RI 02192, U.S.A.}
\bigskip
\centerline{Samir D. Mathur}
\smallskip
\centerline{{\it Department of Physics, The Ohio State University}}
\centerline{\it Columbus, OH. 43210,  U.S.A.}
\bigskip

\medskip

\noindent

\def\gr{g_{rr}(r)}
\def\gphi{g_{\phi\phi}(r)}
\def\loren{(1-g_{rr}(r) {\dot r}^2 -g_{\phi\phi} {\dot \phi}^2)}
\def\dor{{\dot r}}
\def\dophi{{\dot \phi}}

It has been recently suggested that gravitons moving in $AdS_m
\times S^n$ spacetimes along the $S^n$ blow up into spherical
$(n-2)$ branes whose radius increases with increasing angular
momentum. This leads to an upper bound on the angular momentum, thus
``explaining'' the stringy exclusion principle. We show that this
bound is present only for states which saturate a BPS-like condition
involving the energy $E$ and angular momentum $J$, $E \geq J/R$, where
$R$ is the radius of $S^n$.
Restriction of motion to such states lead to a noncommutativity of
the coordinates on $S^n$. As an example of motions which do not obey the
exclusion principle bound, we show that
there are finite action instanton configurations interpolating between
two possible BPS states. We suggest that this is consistent with the
proposal that there is an effective description in terms of
supergravity defined on noncommutative spaces and noncommutativity
arises here because of imposing supersymmetry.

\Date{August, 2000}

\newsec{Introduction and Summary}

One of the striking consequences of the holographic correspondence in
$AdS_m \times S^n$ spacetimes \ref\ads{J. Maldacena, Adv. Math. Theor.
Phys. 2 (1998) 231, hep-th/9711200; S.S. Gubser, I.R. Klebanov and
A.M. Polyakov, Phys. Lett. B 428 (1998) 105, hep-th/9802109;
E. Witten, Adv. Theor. Math. Phys. 2 (1998) 253, hep-th/9802150.}
is the stringy exclusion
principle \ref\exclsuion{J. Maldacena and A. Strominger, JHEP 9812
(1998) 005, hep-th/9804085}.
It has been also argued that the stringy exclusion principle means
that the dual supergravity should live on a noncommutative space-time,
e.g. quantum deformations of $AdS \times S$
\ref\ramjev{A. Jevicki and S. Ramgoolam, JHEP 9904 (1999) 032,
hep-th/9902059; P. Ho, S. Ramgoolam and R. Tatar, Nucl. Phys.
B 573 (2000) 364, hep-th/9907145;
A. Jevicki, M. Mihailescu and S. Ramgoolam,
hep-th/0006239.}. The principle states that the maximum angular
momentum of single particle BPS states (in space-time) is bounded by
$N$, where $N$ is the flux of the $n$-form magnetic field strength on
the sphere. Recently, McGreevy, Susskind and Toumbas
\ref\giant{J. McGreevy, L. Susskind and N. Toumbas, JHEP 0006 (2000)
008, hep-th/0003075.} has provided a novel explanation of this
phenomenon. According to their proposal, a single trace operator of
the holographic theory are well described as single particle
supergravity modes (we call generically call them gravitons) for low
values of the angular momenta. However, for large angular momenta,
these blow up into spherical $(n-2)$ branes moving on the $S^n$.
In some cases this blow-up follows qualitatively from the
Myers effect \ref\myers{R. Myers, JHEP 9912 (1999) 022,
hep-th/9910053; D. Kabat and W. Taylor, hep-th/9711078.} 
\foot{ This connection has been explored in
\ref\dtriv{S.R. Das, S. Trivedi and S. Vaidya,
{\it to appear}}}.
It is then demonstrated that the radius of these spherical branes
grows with the angular momentum for a restricted class of motion where
the radius of the brane does not change with time and in addition
there are no waves along the brane.  Since the radius of the brane
cannot exceed the radius of the sphere on which it moves, there is a
bound on the angular momentum.

As we will see, it is important to emphasize that for consistency
there should be only one BPS state for a given angular momentum even
though it might appear that there is both a graviton and a ``giant
graviton'' or a wrapped brane. The point is that the former is a
valid description for small angular momenta while the latter is the
correct description for large angular momenta.

A natural question arises immediately. What about non-BPS states ?
There are of course a large number of non-BPS states in the gauge
theory which are still represented by single trace operators and
should therefore correspond to single particle or single brane states
in the dual description. In this paper we show that the hamiltonian
for brane motions on spheres leads to a BPS-type bound, viz. there is
a lower bound on the energy for a given angular momentum.  The
restricted type of motion considered in \giant\ are those which
saturate this bound.  If we are working in a supersymmetric theory, it
is natural to expect that these are also configurations which preserve
some of the supersymmetries. Thus non-BPS states correspond to other
motions, e.g. oscillations and changes of the radius.  From this bound
it is straightforward to see that the radius of the brane increases
with the angular momentum for $n > 3$. For such motions the potential
has two minima - one at zero radius and the other at a radius which
scales as $J^{1\over n-3}$. We show that there are finite action
instanton configurations interpolating between the two minima.  The
implication of this is, however, less clear since the description of
the system as a brane with some DBI action fails when the size of the
brane is small.

  Since the dynamics of D-branes is defined on a commutative phase
  space one might wonder whether noncommutativity is visible in the
  ``giant graviton'' scenario. We show that this is in fact true . More
  precisely we show that the restriction of the motion of the branes to
  those which saturate the BPS bound which can be implemented in terms
  of supersymmetry generators imply a set of second class constraints
  in the phase space.  This implies that the Dirac brackets of two
  coordinates on the $S^n$ transverse to the $(n-2)$-brane are nonzero
  which should imply that in the quantum problem these operators do not
  commute. We show that the Dirac bracket in fact becomes singular at
  $n=3$, which is consistent with the fact that in this case all such
  states have the same (maximal) angular momentum. One could arrive at
  the same conclusion by considering directly the quantum commutators
  of the coordinate operators projected on to the subspace of states
  implied by the restriction on brane motion. Thus in the reduced phase
  space, it appears that two space directions do not commute. This is
  quite similar to the problem of a charge particle moving on a 2-plane
  in the presence of a constant mangetic field. In that case,
  restriction to the lowest Landau level implies that the two spacelike
  coordiantes do not commute. Connections of the bound on angular
  momenta of giant gravitons with noncommutativity have been
  heuristically discussed earlier in \ref\holi{P. Ho and M. Li,
  hep-th/0004072}.  However the precise origin of noncommutativity is
  rather unclear in this treatment.

The results of this paper were reported in ``KIAS Summer Workshop
on Branes'' \ref\antseoul{Talk given
by A. Jevicki, KIAS Summer Workshop on Branes'', Seoul, June 12-17, 2000,
http://www.kias.re.kr/conf/brane-program.html}. While this paper was
under preparation the papers \ref\giantnewa{M. Grisaru,
R. Myers and O. Tafjord, hep-th/0008015} and
\ref\giantnewb{ A. Hashimoto, S. Hirano
and N. Itzhaki, hep-th/0008016.} appeared on the net which have some
overlap with Sections 2 and 3 of our work. In Section 4. we use some
results of \giantnewa\ and \giantnewb\ to describe the origin of
noncommutative space in the theory.

\newsec{Brane motion on Spheres}

We consider space-times of the form of $AdS_m \times S^n$ where
$m+n = 10$ in string theory and $m+n = 11$ in M-theory. The radius
of $S^n$ is $R$, which is also the scale of the $AdS$ space-time and
there is a constant $n$-form flux on $S^n$ with $N$ quanta of flux.
Let us consider the sphere $S^n$ embedded in $R^{n+1}$ with
coordinates $X^1 \cdots X^{n+1}$
\eqn\one{(X^1)^2 + \cdots (X^{n+1})^2 = R^2}
and we
choose coordinates on the sphere as follows
\eqn\onea{\eqalign{& X^1 = {\sqrt{R^2 - r^2}}~\cos\phi \cr
& X^2 = {\sqrt{R^2 - r^2}}~\sin\phi }}
where $0 \leq r \leq R$.
The remaining $X^3 \cdots X^{n+1}$ chosen to satisfy
\eqn\two{(X^3)^2 + \cdots (X^{n+1})^2 = r^2}
These may be written in terms of $p = (n-2)$ angles $\theta_1
\cdots \theta_p$ and $r$ in the form of standard spherical
polar coordinates in $p+1$ dimensions. Then the metric on
$S^n$ becomes
\eqn\three{ds^2 = {R^2 \over R^2 - r^2}dr^2
+ (R^2 - r^2) d\phi^2 + r^2 d\Omega_p^2}
where $d\Omega_p^2$ is the volume element on a unit $p$-sphere.

We consider $p$-branes, with $p = n-2$, wrapped on the $S^p$
embedded in $S^n$,
moving entirely in the $S^n$
and sitting at the center of global coordinates in the $AdS_m$.
The time coordinate in $AdS$ is denoted by $t$. In the $(p+1)$
dimensional world volume of the brane with coordinates
$\tau, \sigma_1 \cdots \sigma_p$ we choose a static gauge
\eqn\four{\tau = t~~~~~\sigma_i = \theta_i ~~(i = 1,\cdots p)}.
The dynamical coordinates are now $r (t,\theta_i)$ and
$\phi (t, \theta_i)$.

We will look at motions of the brane where there are no oscillations,
i.e. $r,\phi$ are independent of the angles $\theta_i$. Then the
brane lagrangian is given by
\eqn\four{L = -\lambda[r^p (1-g_{rr}(r) {\dot r}^2
-g_{\phi\phi} {\dot \phi}^2)^{1/2} - r^{p+1} {\dot \phi}]^{1/2}}
where
\eqn\five{\eqalign{& \lambda = {N \over R^{p+1}}\cr
& g_{rr}(r) = {R^2 \over R^2 - r^2}\cr
& g_{\phi\phi}(r) = R^2 - r^2}}
The first term is the Dirac-Born-Infeld (DBI) term. The coefficient
is a rewriting of the tension of the brane in terms of $N$ and $R$.
This follows from the corresponding classical supergravity solution.
It is crucial in what follows that we have exactly the same coefficient
in the second term - the Chern-Simons term. This is the coupling of
the brane with the $n$-form field strength and the precise coefficient
follows from standard flux quantization.

In the following we will set $R=1$ so that all dimensional quantities
are in units of $R$. We will restore $R$ at the very end.

The canonical momenta for $r$ and $\phi$ are $p_r$ and $p_\phi$
respectively and are given by
\eqn\six{\eqalign{&p_r \equiv \lambda P = {\lambda r^p \gr \dor
\over \loren^{1/2}}\cr
&p_\phi \equiv \lambda j = {\lambda r^p \gphi \dophi \over
\loren^{1/2}} + \lambda r^{p+1}}}
The momentum $p_\phi$ is an angular momentum and is conserved. $p_r$
is not conserved.
From \six\ one gets
\eqn\seven{\loren^{1/2} = r^p [r^{2p} + {P^2 \over \gr}
+ {(j-r^{p+1})^2 \over \gphi}]^{-1/2}}

The canonical hamiltonian can be now derived in a standard fashion
and becomes
\eqn\eight{H = \lambda [r^{2p} + {P^2 \over \gr}
+ {(j-r^{p+1})^2 \over \gphi}]^{1/2}}

\subsec{BPS bounds}

Motion can be labelled by the quantum number $j$. It is easy to show that
for some given $j$ there is a lower bound on the energy - a BPS bound.
This is not immediately obvious from the form of the hamiltonian
\eight. However a straightforward algebra allows us to rewrite $H$ in
the following form
\eqn\nine{H = \lambda [j^2 + {P^2 \over \gr} +
{(jr - r^p)^2 \over \gphi}]^{1/2}}
Since $\gr = (1-r^2)^{-1}$ and $\gphi = 1-r^2$ (in $R=1$ units) are
positive it is clear that
\eqn\ten{H \geq \lambda j}
This is the BPS bound.

In deriving the form of the hamiltonian given in \nine\ it is
absolutely crucial that the relative coefficient between the DBI term
and the Chern-Simons term is what it is. This happens because the
$n$-form flux is quantized in the standard way.  Furthermore the exact
form of the metric on the sphere is also crucial. All the details of
working in a consistent supergravity background has entered in the
calculation.

\subsec{BPS saturated states and angular momentum bounds}

The bound is saturated when
\eqn\eleven{ p_r = 0}
and
\eqn\twelve{ jr = r^p}
The latter has two solutions for $p \neq 0$
\eqn\thirteen{ r= r_1 = 0~~~~~~~r = r_2 = j^{{1\over p-1}}}
Thus BPS motions have constant $r$, which is the size of the brane.

The potential energy for such motion
\eqn\thirteena{V(r) = {(jr - r^p)^2 \over 1-r^2}}

For $p=0$ the potential does not vanish at
$r = 0$. There is a minimum for $j < 1$ in the physical range of $r$.
however the potential is nonvanishing at the minimum and this does
not correspond to a BPS state. When $ j > 1$ there is one minimum where
$V(r) = 0$ and the location of the minimum moves to smaller values of
$r$ as $j$ increases. Thus, for $p = 0$ BPS states must have $j > 1$.

For $p = 1$ and $j \neq 1$ there is no minimum for $r < 1$. For $j =1$
the potential is zero everywhere. Thus all BPS states have $j =1$.

For $p \geq 2$ the potential has two minima with a maximum inbetween.
These minima are precisely $r_1$ and $r_2$ given in \thirteen. Thus
there are two kinds of BPS states : the one which correspond to zero
size branes and the other with branes with sizes scaling as $j^{1/(p-1)}$.
Since the range of $r$ is between $0$ and $1$ this immediately implies
that there is an upper bound for $j$
\eqn\fourteen{j \leq R^{p+1}}
where we have restored $R$. The physical angular momentum is
\eqn\fifteen{ p_\phi = N}

For a BPS state, $H = \lambda j = p_\phi/R$. 
This is the {\it same} dispersion
relation as that of a massless graviton which is moving purely on the
sphere. What is surprising is that states of branes, which are by
themselves heavy objects, can lead to a light state. The reason this
behind this is of course the coupling to the $n$-form field strength. The
effect of this cancelled the effect of brane tension.

In the above discussion, we have used the phrase ``BPS configuration'' in
its original sense. In a supersymmetric theory one would expect that
these configurations also preserve some of the supersymmetries \foot{This
has in fact been shown in \giantnewa,\giantnewb.}

\newsec{Tunnelling between vacua}

We saw that for $p \geq 2$ there are two minima. Strictly speaking,
the minimum at $r=0$ is in a regime where we cannot trust our picture.
The description of brane physics in terms of a DBI-CS lagrangian is
valid when the size of the brane is much larger than the basic length
scale of the theory and clearly a zero size brane cannot be described
in this fashion. On the other hand, for sufficiently large $j$ the
other minimum lies in the domain of validity of our calculation. This is
consistent with the overall picture implied in \giant. For low values of
the angular momentum the perturbative graviton is a good description of the
state. For large values of angular momenta, this description fails
and one should consider the states as wrapped branes.

It is nevertheless of some interest to ask whether there are
finite action tunelling configurations between the two vacua. We now
want to consider motion in the $r$ direction, for a given value of
$j$ in euclidean signature.  The hamiltonian for such motion is
given by
\eqn\sixteen{ H = \lambda [ U(r) + {p_r^2 \over \lambda^2 \gr}]^{1/2}}
where
\eqn\sixteena{U(r) = j^2 + V(r)}
The corresponding lagrangian is then
\eqn\seventeen{L = -\lambda [U(r)]^{1/2}[1-\gr~\dor^2]^{1/2}}
so that the euclidean action is
\eqn\eighteen{S_E = \int dt[U(r)]^{1/2}~[1+\gr~\dor^2]^{1/2}}
while the euclidean hamiltonian is
\eqn\nineteen{H_E = - \lambda[U(r) - {p_E^2 \over \lambda^2 \gr}]^{1/2}}
where
\eqn\twenty{p_E = \lambda {\gr~\dor [U(r)]^{1/2}
\over [1 + \gr \dor^2]^{1/2}}}
To construct instanton solutions interpolating between the two minima
of $U(r)$ we need solutions of the euclidean equations of motion with
euclidean energy $H_E^2 = \lambda^2 j^2$. It is easily seen that such
motion obeys
\eqn\twentyone{\dor = \pm {1\over j}r~(j-r^{p-1})}
the two signs corresponding to instantons and anti-instantons. Using
\twentyone\ and \eighteen\ it is easy to check that the euclidean
action for this solution is
\eqn\twentytwo{S'_{ins} = {\lambda \over j}\int dt ~U(r)}
This action may be easily seen to be infinite. However one must remember
that the energy of the states between which tunelling is occuring is
nonzero and equal to $\lambda j$. This has itself an action
\eqn\twentythree{S_0 = \lambda j \int dt}
Thus the true instanton action is
\eqn\twentyfour{S_{ins} = S'_{ins} - S_0}
and this is in fact finite.

For example for $p =2$ the solution to \twentyone\ is
\eqn\zone{ { r \over j-r} = e^t}
and the action $S'_{ins}$ is
\eqn\ztwo{ S'_{ins} = \lambda \int_0^j dr [{j^2 \over r(j-r)}
+ {r(j-r) \over 1-r^2}]}
The first term in the integral is clearly divergent while the
second term is finite. However,
using \twentyone\ again one sees that this term is in fact
\eqn\zthree{ \lambda j \int dt}
which is exactly $S_0$. Thus the subtracted quantity $S_{ins}$ is
finite :
\eqn\twentyfive{S_{ins} = \lambda[j + {1\over 2}(1-j) \log (1-j)
- {1\over 2}(1+j) \log (1+j)]}

We have shown that there are finite action instanton configurations
which interpolate between the two minima of the potential. However,
as emphasized above, the meaning of this is not very clear since the
configuration with zero sized branes clearly lies outside the validity
of our description.
In fact for a given angular momentum there is only one state : for
low angular momentum this is a point like state represented by a graviton
and for large angular momentum this is an extended brane. For intermediate
angular momenta the description is probably complicated.

\newsec{Multiple brane states}

The $N=4$ super Yang Mills theory that arises from D-3-branes has
   chiral operators of the form $\tr[\Phi^{i_1}\dots \Phi^{i_n}]$, where we
   symmetrize in the indices $i_k$.  But it has also been argued that
   there exist multi-trace operators that are also chiral primaries
   \ref\multi{L. Andrianopoli and S.  Ferrara, Lett.Math.Phys. 48
   (1999) 145; W.  Skiba, Phys.Rev. D60 (1999) 105038; M.  Bianchi,
   S. Kovacs, G.Rossi, Yassen and S.  Stanev, JHEP 9908 (1999)
   020.}. These operators are of the form (for two traces)
\eqn\qone{\tr[\Phi^{i_1}\dots
\Phi^{i_m}]\tr[\Phi^{i_{m+1}}\dots \Phi^{i_n}]}
with the indices $i_k$ again symmetrized.  In a similar manner we can
make operators with more traces. These operators are expected to be
dual to multi-particle states in the dual string theory. In the case
of $AdS_3\times S^3\times M^4$ it was found in \ref\deboer{J. de Boer,
Nucl.\ Phys.\ {\bf B548}, 139 (1999) [hep-th/9806104]; JHEP {\bf
9905}, 017 (1999) [hep-th/9812240].} that multiparticle states in
supergravity were needed to account for the elliptic genus computed
from the dual CFT.

The existence of multiparticle chiral primaries raises the following
issue for the stringy exclusion principle. To be able to get these
states we must be able to construct multiparticle states where the
interactions between the particles exactly `cancel out' giving energy
equal to the sum of $R$-charges. At the same time we should not be
able to increase the number of such quanta without bound, since when
the total charge exceeds the limit set by the exclusion principle then
the state should not be BPS.

Let us examine the consequence of this fact for the giant
gravitons. The $n$-trace operators in the gauge theory would
presumably be dual to $n$ branes placed in the dual spacetime. If this
state is to be BPS, then the interactions between these branes must
`cancel out'. This $n$ particle state is different from the single
particle state with the same charges, so we do not require that these
configurations mix to produce one effective state.  (Note however that
operator refinitions mix single and multiparticle states in some cases
\ref\mix{Hong Liu and  A.A. Tseytlin, JHEP 9910
(1999) 003;  E. D'Hoker,  D.Z. Freedman, S.D.
Mathur, A. Matusis and  L. Rastelli, to appear in
the Yuri Golfand Memorial Volume `Many faces of the Superworld'
(ed. M. Shifman).
(hep-th/9908160) }.)

On the other hand when the total charge on the branes exceeds the
limit set by the exclusion principle, we should no longer be able to
make a BPS state. Thus consider the giant graviton that expands in the
$AdS$ direction rather than along the sphere,
and let the angular momentum $L$ exceed the limit set
by the exclusion principle. Then there can be a tunneling from this
state to one where there are say two giant gravitons, with angular
momenta $L_1, L-L_1$.  Pictorially, we imagine a large sphere pinching
in the middle and separating into two spheres.

While we have not computed the action for such an instanton, there
appears to be no reason why it should diverge. In the case of
tunneling from a finite brane to a point, the latter configuration was
singular and one could worry about corresponding divergences in the
amplitude. But the tunneling on hand is between two regular
configurations, and we can make interpolating configurations that have
finite contribution from the Born-Infeld and Wess-Zumino terms.

Thus the picture of tunneling may be more complex than that noted in
\giantnewa\ and \giantnewb.  Apart from the pointlike graviton and
the two giant gravitons, we have a host of multi-brane states that
have the same quantum numbers. The exclusion principle requires some
of these to exist, while the others (with $L$ larger than the
exclusion bound) may disappear by tunnelings that involve all the
configurations mentioned above.

\newsec{Supersymmetry and Noncommutativity}

We have seen that for BPS motions wrapped branes have a bound on the
angular momentum thus providing a new perspective on the stringy
exclusion principle.

Bounds on angular momentum appear naturally for particle motions on
noncommutative spaces, e.g. fuzzy spheres or quantum spheres. One
might wonder : does the dynamics discussed above which is entirely
based on a commutative space imply an (effective) dynamics in a
noncommutative space ?

The important point here is that it is only for BPS states there is
a bound on the angular momentum, not for other motions like changes of
the size of the brane or oscillations of the brane. Likewise from the
point of view of holography there are non-BPS states which can have
any value of the angular momentum. Consider for example $AdS_5 \times
S^5$ where the holographic theory is $3+1$ dimensional $N=4$ Yang-Mills
theory. In terms of the Higgs fields $\Phi^i~i = 1,\cdots 6$ of this
theory chiral primary operators are of the form
\eqn\bone{{\rm Tr}_S [\Phi^{i_1}\cdots \Phi^{i_n}]}
where the subscript ``S''
means that we have to symmetrize with respect to the
indices $i_1 \cdots i_n$ and subtract the trace. Supergravity modes
lie in the chiral primary multiplet obtained from \bone\ by acting with
supersymmetry charges. Clearly \bone\ has a $SO(6)$ angular momentum
equal to $n$. The rank $N$ of the gauge group $SU(N)$ is in fact the
quantized flux of the five form field strength on $S^5$. Such operators
and their supersymmetry partners can have a maximum angular momentum $N$.
There are however operators which involve derivatives of $\Phi$
which do not create BPS states, e.g.
\eqn\btwo{ {\rm Tr}_S [\Phi^{i_1}\partial \Phi^{i_2}
\partial \Phi^{i_3} \cdots \Phi^{i_n}]}
Clearly there is no bound for the angular momenta of these operators
since we can have higher and higher derivatives and $\partial^n \Phi$ is
a different matrix than $\Phi$.

Knowing that BPS states respect half of the supersymmetries
we are now lead impose the condition $Q = 0$ on the phase space .
  This can be done since
$Q$ represent  symmetry generators of the theory.The imposition
of this symmetry strongly on the phase space will lead ,as we will now
argue to a noncommutative space.

We therefore need to know what kind of motion of the brane respect
half the supersymmetries of the background. The question we ask is
the converse of the question answered in \giantnewa\ and \giantnewb,
where it was shown that the giant graviton with no motion in the $r$
direction respects half the supersymmetries. On the worldvolume action
of the brane with fermionic coordinates $\Theta^\alpha$ we need to
choose a $\kappa$- symmetry gauge condition $(1 + \Gamma)\cdot \Theta
= 0$ where $\Gamma$ is is the pullback
\eqn\cone{\Gamma = {1\over (p+1) !}\epsilon^{i_1 \cdots i_{p+1}}
\partial_{i_1}X^{\mu_1} \cdots \partial_{i_{p+1}}X^{\mu_{p+1}}
\Gamma_{\mu_1 \cdots \mu_{p+1}}}
where $X^\mu$ denote the bosonic coordinates on the brane worldvolume,
$\Gamma_\mu$ are the Dirac Gamma matrices in target space and
$\Gamma_{\mu_1 \cdots \mu_n} = \Gamma_{\mu_1}\cdots \Gamma_{\mu_n}$
We need to look at supersymmetry transformations which preserve this
gauge choice. These are the transformations
\eqn\cthree{\delta \Theta = {1\over 2} ( 1- \Gamma)\epsilon}
for an infinitesimal spinor parameters $\epsilon$. To preserve the
supersymmetries of the background we require in addition that $\epsilon$
is in fact a Killing spinor of the background. The corresponding
supercharge is given by
\eqn\cfour{ Q  = {1\over 2} {\bar \Theta}\cdot
(1-\Gamma)}
so that $\delta \Theta = \{ Q \cdot \epsilon, \Theta \}$.

In the static gauge we
are using the expression for $\Gamma$ may be seen to be (using
expressions given in \giantnewb)
\eqn\cfive{\eqalign{\Gamma =
{1\over r^{p+1}}[{H \over \lambda}\Gamma_0 &
+ {P}({\sqrt{1-r^2}}\sin\phi \Gamma_{p+2} + {\sqrt{1-r^2}}\cos\phi
\Gamma_{p+1})\cr
& + (j-r^{p+1})
({\cos\phi\over {\sqrt{1-r^2}}} \Gamma_{p+2} -
{\sin\phi\over {\sqrt{1-r^2}}} \Gamma_{p+1})] \Gamma_{p\cdots 1}}}
where we have now used a coordinate system on the sphere $S^{p+2}$
which have angles $\theta_1 \cdots \theta_{p+2}$. The angles $\theta_1
\cdots \theta_p$ are on the $S^p$ on which the $p$-brane is wrapped
while the relationshipe between $\theta_{p+1}, \theta_{p+2}$ with
$r,\phi$ are given by
\eqn\csix{{\sqrt{1-r^2}}\cos\phi = \cos \theta_{p+2}
~~~~~~~~~{\sqrt{1-r^2}}\sin\phi = \sin\theta_{p+2} \cos\theta_{p+1}}

Using an analysis similar to that in \giantnewa\ and
\giantnewb\
it is straightforward to see from \cfive\ and \cfour\ that the
condition that half of the supercharges vanish 
implies $P = {p_r \over \lambda} = 0$.

What does this condition imply in phase space ?
The coordinates $(r,\phi)$ and the momenta $(p_r,p_\phi)$
satisfy standard Poisson brackets. In particular
\eqn\bzero{ [r,\phi]_{PB} = 0}
However, we have shown that BPS motions have $p_r = 0$.
Thus we should regard this as a constraint
in phase space
\eqn\btwo{ \psi_1 = p_r = 0}
as a weak condition. Taking Poisson bracket with the Hamiltonian
yields a secondary constraint
\eqn\bthree{ \psi_2 = {d V \over dr} = 0}
Actually, this can also be seen as following from the symmetry reduction
condition $Q = 0$. There are no further constraints. This is because a direct
computation
yields
\eqn\bfour{ [H,\psi_2]_{PB} = -{\lambda^2 p_r \over H~\gr}
{d^2 V \over dr^2}}
which vanishes on the constraint surface because of \btwo. Finally
the two constraints $\psi_1$ and $\psi_2$ form a second class system
with the Poisson bracket
\eqn\bfive{ [\psi_1, \psi_2 ]_{PB} = - {d^2 V \over dr^2}}
so that the matrix of PB's of the constraints is
\eqn\bfivea{C = -i {d^2 V \over dr^2} \sigma_2}
where $\sigma_2$ is the Pauli matrix

To analyze the dynamics of these restricted set of motions we need
to look at the brackets of unconstrained variables
on the reduced phase space. Alternatively we should look at Dirac
brackets. These may be computed in a straightforward manner and
the result is
\eqn\bsix{ \eqalign{ [r,\phi]_{DB} & = 
[r,\phi]_{PB} - [r,\psi_1]_{PB} (C^{-1})^{12}
[\psi_2, \phi]_{PB} \cr
& = - {1 \over \lambda} {\partial^2 V / \partial r \partial j
\over \partial^2 V / \partial r^2}}}
Onn the constraint surface this evaluates to
\eqn\bseven{ [r,\phi]_{DB} = {R^{p-1} \over N} {r^{2-p}
\over p-1}}
which is nonzero. In a quantum theory we should replace this Dirac
bracket by a commutator and one would have noncommuting coordinates
on the sphere. The noncommutativity is proportional to $1/N$ as expected.
Furthermore this is divergent for $p = 1$ and reverses sign for $p = 0$.
However these are the two cases where is no true bound for the
angular momentum.

Alternatively one can consider the quantum theory directly. Now the
condition $p_r = 0$ should be imposed on the space of states. If
$P$ denotes the projection operator on these states the relevant
dynamical quantities in this subspace of states are $P r P$ and
$P \phi P$. These will not commute even though $r$ and $\phi$ do.

The origin of noncommutativity in our problem is similar to the way
noncommutativity arises in the quantum Hall effect when one restricts
to the lowest Landau level. Here again the restriction to the lowest
Landau level may be viewed at the classical level as constraints
which set the velocities to zero as weak conditions. The Dirac bracket
for the coordinates is then nontrivial.One justification of this procedure
for restricting to lowest Landau level is given by taking the mass zero limit.
In the present case we have a much more elegant reason for such
reduction namely supersymmetry.

Interestingly, the above commutation relation turns out to be identical
to the one obtained by the following heuristic argument in \holi\
where it is also shown that for $p = 2$ these
are the same as the commutators which define a fuzzy $S^4$.
We have seen that for motions with $P = 0$ the size of the brane is
related to the angular momentum $j$ by the relation $j = r^{p-1}$.
We can then consider $r^{p-1}$ as the canonical conjugate to $\phi$.
This leads to a commutation relation between $r$ and $\phi$ which
is the same obtained by replacing the Dirac bracket \bseven\ by
a commutator. However this heuristic argument does not throw light
on the origin of noncommutativity which lies in the fact that we
are working on a subspace of states.Most importantly as we have
argued the reduction to a noncommutative space can be understood as a
hamiltonian reduction based on supersymmetry.

\newsec{Conclusions}

Our analysis has provided an important consistency check on the
giant graviton picture viz. BPS states have bounded angular momenta
while there are non-BPS states which can have arbitrary angular
momenta. This is consistent with the stringy exclusion principle.
We have shown that at the classical level such states have the
same dispersion relation as that of a graviton ; the brane tension
is cancelled by the Lorentz force due to the field strength to
which the brane couples. Furthemore the very existence of the BPS bound
required precise coefficients in front the DBI and Chern Simons terms
- these incorporate flux quantization as well as the details of the
geometry.

In \giantnewa\  and \giantnewb\ it has been shown that these BPS
states are in fact those which preserve half of the supersymmetries
of the background. We showed that one can impose these supersymmetry
conditions only when $p_r = 0$ or equivalently the size of the
brane is fixed during its motion. We can then impose the condition
that half of the supercharges vanish
as a strong condition which would then imply this
restriction of the motion of the branes which
which lead to bounded angular momenta for $p \geq 2$.
The same restriction also led us
to the fact that for such motions, the two transverse coordinates
on the sphere can be regarded as noncommuting. It is important to
realize that there is nothing noncommuting at the fundamental level.
This arises purely because we are considering motion on a constrained
surface on phase space. This constrained surface can be understood as
given by the condition that on it half of the supercharges vanish. 
In other words space can be regarded as
noncommutative if and only if we restrict to the subspace of BPS states.
We believe that this fact can have implications for the suggestion
that noncommutativity could be the origin of the stringy exclusion
principle.

\newsec{Acknowledgements}

We would like to thank J. Hoppe, S. Ramgoolam and S. Trivedi for
discussions.  This work was partialy supported by Department of Energy
under grant DE-FG02-91ER4068-Task A. S.D.M. is supported in part by
DOE grant no. DE-FG02-91ER40690.
S.R.D. would like to thank the
String Theory Group at Harvard University, Department of Physics, Ohio
State University and Department of Physics at Brown University for
hospitality. 

\listrefs
\end